\documentclass{article}

\input{tcilatex}

\begin{document}

\textbf{TOWARDS CONSCIOUS STOCHASTIC SYSTEMS}

\bigskip

E. A. Novikov

\bigskip Institute for Nonlinear Science, University of California - San
Diego, La Jolla, CA 92093 - 0402

\bigskip

A modeling procedure for enhancing performance of stochastic systems is
proposed.

\bigskip

A natural, but expensive, approach to enhance performance of systems is to
construct a hierarchy of observations and control. On the first level of
this hierarchy systems perform certain functions in response to external
(sensory) input. On the second level systems have ability to observe their
performance and to make corrections, if necessary, according to certain
criteria. On the third level systems have ability to observe the second
level procedures of observation-control and correct them. Presently the cost
of the third and higher level artificial systems seems prohibitive. However,
theoretically, we can imagine a limit in this hierarchy and can call the
resulting systems conscious.

Hierarchical structures appear naturally in systems with strong interaction
of many degrees of freedom. Typical signatures of such hierarchy are
so-called similarity laws. Particularly, in turbulence the concept of
scale-similarity was developed and was associated with the
infinitely-divisible distributions [1]. The activity of the human brain also
revealed the regime of scale-similarity, which was discovered by using the
multi-channel MEG (magnetoencephalogram) [2,3] and EEG
(electroencephalogram)[4]. Hundreds of billions of interconnected neurons
and surrounding sells (particularly, astroglia), apparently, is favorable
playground for hierarchical structures in the brain.

The electrochemical brain activity is taking place in wet and warm
surroundings. To reproduce such activity in artificial systems, even
approximately, seems impossible. However, modeling of the effects of
consciousness [5-9] can be used to enhance performance of systems.

Consider first level system represented by equations:

\begin{equation}
\frac{dx_{k}}{dt}=f_{k}(\mathbf{x,s)},\;k=1,...,n  \tag{1}
\end{equation}

Here $\mathbf{x}(t)=(x_{1},...,x_{n})$ are variables of reaction to the
sensory input $\mathbf{s}(t)=(s_{1},...,s_{m})$, $f_{k}$ are some functions.
Sensory input $\mathbf{s}(t)$, generally, is $m$-dimensional stochastic
process with mean values $<\mathbf{s}(t)>$ and fluctuations $\mathbf{s}%
^{\prime }(t)=\mathbf{s}(t)-<\mathbf{s}(t)>$. Certain probability
distribution is assumed for fluctuations of sensory input. It could be
Gaussian or some empirically found distribution.

Now, in accord with [5-9], we introduce complex variables $\mathbf{z=x+iy}$,
where $\mathbf{y}=(y_{1},...,y_{n})$ are internal variables representing the
effects of consciousness. Substitution of $\mathbf{z}$ instead of $\mathbf{x}
$ \ into (1) gives:%
\begin{equation}
\frac{dx_{k}}{dt}=\func{Re}\{f_{k}(\mathbf{x+iy,s})\}  \tag{2}
\end{equation}

\begin{equation}
\frac{dy_{k}}{dt}=\func{Im}\{f_{k}(\mathbf{x+iy,s})\}  \tag{3}
\end{equation}

Equations (2) and (3) are coupled, assuming that at least some of functions $%
f_{k}$ are nonlinear relative to $\mathbf{x}$. Initial values $y_{k}(0)$ can
play a role of adjustable parameters (see below). The effect of such
procedure, naturally, depends on functions $f_{k}$. If all these functions
are linear relative to $\mathbf{x}$, there will be no effect: equation (2)
will be reduced to (1) and equation (3) will be irrelevant to the
performance of system. Nonlinearity is essential to this approach (in Ref.
5-9 nonlinearity is determined by the sigmoidal firing rate of neurons). In
more general procedure quaternions can be used instead of complex variables.
In Ref. 6-9 imaginary components of the quaternion correspond to subjective
experiences divided in three groups: sensations, emotions and reflections.

Equations (2-3) and analogous quaternion equation can be considered as an
stochastic attractor of high level system. It will be interesting in future
to prove corresponding theorem with certain restriction on
observation-control and on ($\mathbf{f,s}$). Initial values $\mathbf{y}(0)$
play a role of genetic parameters. In some areas of these parameters the
system (with a given statistics of $\mathbf{s}$) may perform better then in
others, in some it may not work at all.

Note, that complex fields have been used recently [10] to eliminate
classical electromagnetic divergencies, namely, the infinite self-energy of
electrons and the paradoxical self-acceleration. The same (algebraic)
approach works for the quantum interaction of charges. Possible connection
may exist between these fields and fields in the modeling of the effect of
consciousness on the electric currents in the human brain [5-9]. Even more
general connection is suspected in the context of a new type of quantum
cosmology (no "strings" attached) [11-14]. In new interpretation of quantum
theory [15] imaginary trajectories and corresponding impulses play an
important role.

These connections are in accord with the tendency of Nature to use the same
trick in many circumstances. It seems natural to use this trick to enhance
performance of stochastic systems.

\bigskip

\bigskip

\textbf{REFERENCES}

\bigskip

[1] E. A. Novikov, Infinitely divisible distributions in turbulence, Phys.
Rev. E \textbf{50}(5), R3303 (1994)

[2] E. Novikov, A. Novikov, D. Shannahof-Khalsa, B. Schwartz, and J. Wright,
Scale-similar activity in the brain, Phys. Rev. E \textbf{56}(3), R2387
(1997)

[3] E. Novikov, A. Novikov, D. Shannahof-Khalsa, B. Schwartz, and J. Wright,
Similarity regime in the brain activity, Appl. Nonl. Dyn. \& Stoch. Systems
(ed. J. Kadtke \& A. Bulsara), p. 299, Amer. Inst. Phys., N. Y., 1997

[4] W. J. Freeman, L. J. Rogers, M. D. Holms, D. L. Silbergelt, Spatial
spectral analysis of human electrocorticograms including alpha and gamma
bands, J. Neurosci. Meth. \textbf{95}, 111 (2000)

[5] E. A. Novikov, Towards modeling of consciousness, arXiv:nlin.PS/0309043

[6] E. A. Novikov, Quaternion dynamics of the brain, arXiv:nlin.PS/0311047

[7] E. A. Novikov, Manipulating consciousness, arXiv:nlin.PS/0403054

[8] E. A. Novikov, Modeling of consciousness, Chaos, Solitons \& Fractals, 
\textbf{25}, 1 (2005)

[9] E. A. Novikov, Imaginary Fields, arXiv:nlin.PS/0502028

[10] E. A. Novikov, Algebras of charges, arXiv:nlin.PS/0509029

[11] E. A. Novikov, Dynamics of distributed sources, Phys. of Fluids, 
\textbf{15}(9), L65 (2003)

[12] E. A. Novikov, Distributed sources, accelerated universe, consciousness
and quantum entanglement, arXiv:nlin.PS/0511040

[13] E. A. Novikov, Vacuum response to cosmic stretching: accelerated
universe and prevention of singularity, arXiv:nlin.PS/0608050

[14] S. G. Chefranov \& E. A. Novikov, Hydrodynamic vacuum sources of
self-germinative dark energy in accelerated universe without "Big Bang", to
appear in JETP, 2010.

[15] E. A. Novikov, Random shooting of entangled particles in vacuum,
arXiv:0707.3299

\end{document}